\documentclass[preprint2]{aastex}

\shorttitle{Astrophysical Quantities of Cepheids}
\shortauthors{Nordgren et al.}

\begin{document}

\title{Astrophysical Quantities of Cepheid
Variables Measured with the NPOI\\
{\it ApJ in press, vol. 543 no. 2, November 10, 2000}}


\author{Tyler E. Nordgren\altaffilmark{1}, J. T. Armstrong\altaffilmark{2},
M. E. Germain\altaffilmark{1}, R. B. Hindsley\altaffilmark{2},
Arsen R. Hajian\altaffilmark{3}, J.~J.~Sudol\altaffilmark{1},
C.~A.~Hummel\altaffilmark{3}}

\altaffiltext{1}{U.S. Naval Observatory, Astrometry Department, NPOI,
P.O. Box 1149 Flagstaff, AZ 86002-1149; nordgren@nofs.navy.mil;
meg@sextans.lowell.edu; jsudol@sextans.lowell.edu}

\altaffiltext{2}{Remote Sensing Division, Naval Research Laboratory,
Code 7210, Washington, DC 20375; tarmstr@fornax.usno.navy.mil;
hindsley@rsd.nrl.navy.mil}

\altaffiltext{3}{U.S. Naval Observatory, Astrometry Department,
3450 Massachusetts Avenue NW, Washington, DC 20392; hajian@fornax.usno.navy.mil,
cah@fornax.usno.navy.mil}

\begin{abstract}

We present mean angular diameters for two cepheid variables,
$\alpha$ Ursae Minoris and $\zeta$ Geminorum,
determined with the Navy Prototype Optical Interferometer
(NPOI). We present linear radii for these cepheids and
two additional cepheids, $\delta$ Cephei and $\eta$ Aquilae,
previously observed at the NPOI. We find the limb-darkened
angular diameter of $\alpha$ Ursae Minoris and of $\zeta$ Geminorum
to be 3.28 $\pm$ 0.02 and 1.55 $\pm$ 0.09 milliarcseconds
respectively.
Using trigonometric parallaxes, we find the linear
radii of $\alpha$ Ursae Minoris, $\zeta$ Geminorum,
$\delta$ Cephei and $\eta$ Aquilae to be 46 ($\pm$ 3) R$_{\odot}$,
60 (+25, -14) R$_{\odot}$, 45 (+8, -6) R$_{\odot}$, and 69 (+28, -15) R$_{\odot}$
respectively. We compare the pulsation periods and linear radii of
this sample of cepheids, which range in period from three to 11 days,
to theoretical and empirical period-radius and period-radius-mass
relations found in the literature.
We find that the observed diameter of $\alpha$ Ursae Minoris is in excellent
agreement with the predicted diameter as determined from 
both surface brightness techniques and theory only if $\alpha$ Ursae Minoris
is a first overtone pulsator.

\end{abstract}

\keywords{Cepheids --- stars: fundamental parameters --- stars: individual
(Polaris, $\zeta$ Geminorum, $\delta$ Cephei, $\eta$ Aquilae)}

\section{INTRODUCTION}

Accurate stellar radii are important
for the study of cepheid
mass, pulsation and distance.
Direct radius measurements of bright,
nearby cepheids allows for comparison to radii found by
indirect and/or theoretical
methods such as numerical models
\citep{bon98}, the infrared flux method \citep{fly89} and
surface brightness relations \citep{mab87,las95}. These
methods are easily applied to distant cepheids
including those in nearby galaxies \citep{gie00}.
Each of these indirect methods results in
period-radius and period-radius-mass relations
which yield different
radii, and different masses, at very small and very large periods.
From directly measured radii we may make comparisons
with these relations.
Since there will always be cepheids too small or too faint
for direct measurement, the comparison between these
indirect measurements and relations is crucial for
the radius estimation of ever more distant cepheids.

At its current magnitude limit (m$_V$ $\sim$5) and longest baseline
(38 meters) the angular diameter of four cepheids are measurable with the NPOI:
$\delta$ Cephei, $\zeta$ Geminorum, $\eta$ Aquilae and
$\alpha$~Ursae Minoris (hereafter Polaris).
In this paper we present
mean angular diameters and compare linear radii for
all four cepheids with those in the literature and with published period-radius,
period-mass-radius, and period-mass relations.
Even though the sample is small, these four cepheids span an interesting
range in pulsation period and characteristics.

\section{OBSERVATIONS AND DIAMETER MEASUREMENTS}

Polaris and $\zeta$ Gem
were observed over the course of two years.
Polaris was observed on 10 nights from
September to November 1997, while $\zeta$ Gem was observed for
four nights: 
12 October 1998, and 20, 23 and 24 February 1999.
The detailed observing strategy and data reduction techniques for
obtaining mean angular diameters at the NPOI
are described in \citet{ten99}. \citet{arm00} present the specific
observations
and data reduction of $\delta$ Cep and $\eta$ Aql. Comparisons
between the reduction method employed by \citet{arm00} and
that used in this work are made at the end of this section.

Briefly, as described in \citet{ten99}
squared-visibilities are measured in each of 10
spectral channels, spaced
evenly in wavenumber, ranging from 649 nm to 849 nm.
A uniform-disk model is fit to the visibility data
from which a uniform-disk diameter is derived.
The uniform-disk diameters of Polaris and $\zeta$ Gem are
found to be
3.14 $\pm$ 0.02 mas and 1.48 $\pm$ 0.08 mas respectively.
As reported in \citet{ten99} the uniform-disk diameters
for $\delta$ Cep and $\eta$ Aql are 1.46 $\pm$ 0.02 mas
and 1.65 $\pm$ 0.04 mas respectively. 
Figure 1 shows visibility data for the NPOI's longest
baseline (East-West) for each of the four cepheids. The
data shown in Figure 1 (a) - (d) are for the night listed
in each. The mean uniform-disk diameter for each is the 
overall mean diameter determined for that cepheid.

\begin{figure}
\figurenum{1}
\plotone{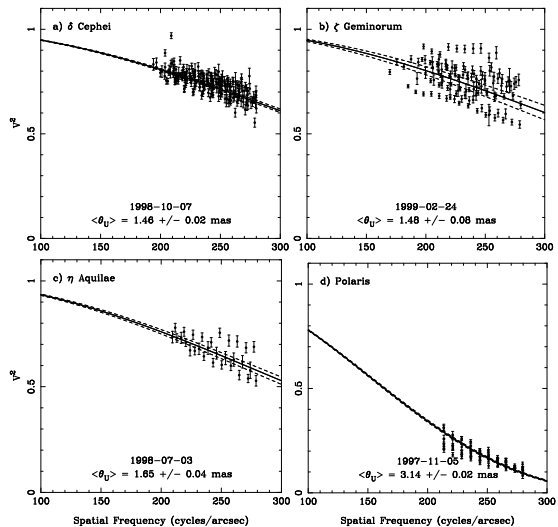}
\figcaption[fig1.ps]{Visibility data for the four cepheids
observed with the NPOI. For each figure, visibilities for
the East-West baseline are shown for a given night. The model uniform
disk diameter and error shown is the mean diameter and error
of the mean for all nights that cepheid was observed. Each
figure is plotted to the same scale. (a) $\delta$ Cephei
for 7 Ocotber, 1998; (b) $\zeta$ Geminorum for 24 February, 1999; 
(c) $\eta$ Aquilae for 3 July, 1998; (d) Polaris for 5 November, 1997.}
\end{figure}

Although limb-darkening of evolved stars has
been directly observed with the NPOI \citep{haj98},
those stars are three times larger than the cepheids
in this study. At the spatial frequencies currently available to
the NPOI the visibility differences between limb-darkened
and uniform-disks for such small stars is less than the scatter
in the data. Until the availability of longer baselines,
limb-darkened
diameters, $\theta_L$, can be derived from
uniform-disk diameters
using a multiplicative conversion factor.
This conversion factor is a single quadratic coefficient
from \citet{cla95} interpolated for the cepheid's average
specific gravity ($\log g$), average effective
temperature, and for the mean central
wavelength of the NPOI bandwidth (740 nm).
The Bright Star Catalogue \citep{hof82} categorizes both
$\zeta$ Gem and Polaris as spectral type F7Ib. For this spectral
type \citet{str81} give a
$\log g$ of 1.71 and an effective temperature of 6000 K.
Using these values and the technique described in \cite{ten99} we derive
a limb-darkened conversion factor (ratio of limb-darkened diameter to
uniform-disk diameter) of
1.046 for both cepheids at 740 nm. The uncertainty in this
conversion factor is estimated to be on the order of 0.5\% \citep{ten99}
even for zeta Gem, whose spectral type is very uncertain.
With this derived limb-darkening coefficient we find
a limb-darkened diameter of
3.28 $\pm$ 0.02 mas for Polaris and 1.55 $\pm$ 0.09 mas
for $\zeta$ Gem.

Using this method in \citet{ten99} resulted in a
limb-darkened diameter of 1.52 $\pm$ 0.02 mas for
$\delta$ Cep and 1.65 $\pm$ 0.04 mas for $\eta$ Aql.
\citet{arm00} use a different reduction method
for the calibration of the raw visibility data for
these two cepheids (as well as two non-variable
``check-stars''). In addition
limb-darkened diameters are fit directly
to the squared-visibility data without first calculating
uniform-disk diameters.
\citet{arm00} measure a limb-darkened angular diameter of
1.520 $\pm$ 0.014 mas for $\delta$ Cep and 1.69 $\pm$ 0.04 mas
for $\eta$ Aql. For the non-variable
star $\beta$ Lac, \citet{arm00} derive a limb-darkened diameter
of 1.909 $\pm$ 0.011 mas while \citet{ten99} finds
1.92 $\pm$ 0.02 mas.
Since these two different reduction methods produced diameters equal
within the errors there is strong confidence in the robustness
of the final results. The diameters for $\delta$ Cep
and $\eta$ Aql used throughout the rest of this work are
those of \citet{arm00}.

Finally, each of the four cepheids is part of
a multiple system.
If the NPOI should detect light from
more than one star, the visibilities measured will be depressed
depending upon the position angle and separation of
the system. If not taken into account this variation
will have the effect of changing the model diameter
that best fits the observed data. Fortunately,
each of the companions is either several magnitudes fainter 
than the cepheid being observed (placing it
well below the
NPOI's detection threshold), at a large enough separation 
($\geq$18 arcseconds) that it
is outside the NPOI's photometric field of view, or both.
For example, the companion to $\eta$ Aql
is 4.6 magnitudes fainter \citep{vap85} while at the
same time being substantially bluer (spectral type
A1V compared to F6Ib-G4Ib for $\eta$ Aql). Since
the visibilities from only the ten reddest channels are used
to fit diameters, further chances of contamination by
the companion are reduced. There is therefore
no indication that there has been contamination
of the measured diameters for any of the cepheids due to stellar companions.

\subsection{DISTANCES AND LINEAR RADII}

Where there is a measured trigonometric parallax, $\pi$, the distance, $d$, to
the cepheid is the reciprocal of $\pi$, and the linear radius is simply:
$R = d \tan(\theta_L/2).$
All four cepheids have parallaxes measured by
Hipparcos \citep{per97},
while $\delta$ Cep and $\eta$ Aql have additional
parallaxes measured at
the U.S. Naval Observatory Flagstaff
Station (H. Harris, 1999, personal communication). For these
two cepheids, the distance and linear radius, are derived from
the weighted mean of the two measured parallaxes: 3.60 $\pm$ 0.53 mas
for $\delta$ Cep and 2.62 $\pm$ 0.74 mas for $\eta$ Aql.
These linear radii are
nearly model-independent; what dependence there is
enters from
the conversion between uniform-disk
and limb-darkened angular diameters, and as previously noted,
is estimated to be at the
level of $\sim$0.5\% of the mean radius.
When the $\sim$70 meter baseline at the NPOI becomes operational
spatial frequencies of $\sim$400 cycles
per arcsecond will
be accessible and at that time limb-darkened angular diameters for
these stars can
be measured directly.

\begin{deluxetable}{lrcccll}
\footnotesize
\tablecaption{Cepheid Angular Diameters, Distances and Radii \label{tbl-1}}
\tablewidth{0pt}
\tablehead{
\colhead{Cepheid} & \colhead{Period\tablenotemark{a}}   &
\colhead{$\theta_{L}$\tablenotemark{b}} & \colhead{$\pi_H$} & \colhead{$\pi_N$} &
\colhead{d\tablenotemark{c}}  & \colhead{$R_N$} \\
\colhead{} & \colhead{(days)}   &
\colhead{(mas)} & \colhead{(mas)} & \colhead{(mas)} &
\colhead{(pc)}  & \colhead{(R$_{\odot}$)}
}
\startdata
Polaris  & 3.9729 & 3.28 $\pm$ 0.02 & 7.56 $\pm$ 0.48 & ... & 132$^{+9}_{-8}$ & 46$^{+3}_{-3}$ \\
$\delta$ Cep  & 5.3663 & 1.52 $\pm$ 0.01 & 3.32 $\pm$ 0.58 & 5.0 $\pm$ 1.3 & 278$^{+48}_{-36}$ & 45$^{+8}_{-6}$ \\
$\eta$ Aql    & 7.1766 & 1.69 $\pm$ 0.04 & 2.78 $\pm$ 0.91 & 2.3 $\pm$ 1.3 & 382$^{+150}_{-84}$ & 69$^{+28}_{-15}$ \\ 
$\zeta$ Gem   &10.1507 & 1.55 $\pm$ 0.09 & 2.79 $\pm$ 0.81 & ... & 358$^{+147}_{-81}$ & 60$^{+25}_{-14}$ \\
\enddata

\tablenotetext{a}{Period for Polaris and $\zeta$ Gem from Don Fernie (private
communication 1999). Period for $\delta$ Cep and $\eta$ Aql from Szabados 1980.}
\tablenotetext{b}{For $\delta$ Cep and $\eta$ Aql from Armstrong et al. 2000.}
\tablenotetext{c}{For $\delta$ Cep and $\eta$ Aql: calculated from the weighted
average of the Hipparcos and USNO parallaxes.}
\end{deluxetable}

Table 1 lists the four cepheids and includes the NPOI limb-darkened
angular diameter, the Hipparcos
parallax, $\pi_H$, USNO parallax, $\pi_N$, the
distance (found from the weighted mean parallax for $\delta$ Cep and
$\eta$ Aql),
and the NPOI's direct linear radius, $R_N$.

\section{DIRECT RADIUS COMPARISONS}


The most common method for estimating
cepheid radii and distances is the Baade-Wesselink,
\citep{wes69} or surface brightness method \citep{bae76}.
This method, of which there are several variations, relies upon
observations of color and radial velocity changes
\citep{mab87}.
For $\zeta$ Gem, \citet{mab87} derive a radius of
65 $\pm$ 12 R$_{\odot}$ while \citet{sca76}
derives a radius of 68 $\pm$ 3 R$_{\odot}$.
The average difference between these and the NPOI
result (Table 1) is 10\% whereas the error of $R_N$
towards higher values is 42\%. The difference between the percent error
towards lower and higher values arises from unequal error bars for
the distance and linear radius in columns 6 and 7 of Table 1.
Similarly for $\delta$ Cep, \citet{mab87}
and \citet{fly89} derive a radius of 41 $\pm$ 2 and 37 $\pm$ 4 R$_{\odot}$
respectively. The average percent difference between $R_N$
and these is 12\% which is slightly smaller than the percent error
toward lower values
of the NPOI radius (14\%). Given the uncertainties
in $R_N$, the measured radii for both these cepheids
are consistent with values in the literature.

\citet{mab87} derive a radius of 55 $\pm$ 4 R$_{\odot}$
for $\eta$ Aql, while \citet{fly89} calculate 53 $\pm$ 5 R$_{\odot}$.
The average percent
difference with $R_N$ is 24\%, slightly larger than the percent error
towards lower values of the NPOI
radius (22\%).
\citet{sas90}, however, use high-resolution
infrared spectroscopy to
find a radius of 63 $\pm$ 6 R$_{\odot}$ for $\eta$ Aql,
almost 10\% larger than those found using optical
spectroscopy.
The photospheric lines in the high-resolution 
infrared spectra show asymmetries and line splitting
which are interpreted to be pulsationally
driven shock waves in the atmosphere. The larger radius results from
the new interpretation of these spectra and projection factors
derived from them specifically for the 
degree of limb-darkening expected in the infrared. The
difference between the diameter derived from the IR spectra and the
NPOI diameter (Table 1) is only 9\%.

Table 2 lists these previously published
radii for three of the cepheids in this paper (not including Polaris). 
The method which shows the least agreement with the
observations reported here is the CORS method,
a variation of the surface
brightness technique which is different in its mathematical
computation \citep{rip97,cor81}. Table 2 shows that for these three cepheids
the CORS method produces radii consistently larger than those produced
by other optical surface brightness methods.

\begin{deluxetable}{lllccccc}
\footnotesize
\tablecaption{Cepheid Radii Comparison\label{tbl-2}}
\tablewidth{0pt}
\tablehead{
\colhead{Cepheid} & \colhead{$R_{N}$}   & \colhead{$R_{SB}$\tablenotemark{a}}   &
\colhead{$R_{IRFM}$\tablenotemark{b}}  &
\colhead{$R_{CORS}$\tablenotemark{c}} & \colhead{$R_{IR}$\tablenotemark{d}}
& \colhead{$R_{BW}$\tablenotemark{e}}
}
\startdata
$\delta$ Cep  & 45$^{+8}_{-6}$  & 41 $\pm$ 2 & 37 $\pm$ 4 & 53 $\pm$ 3 & ... & ...\\
$\eta$ Aql    & 69$^{+28}_{-15}$ & 55 $\pm$ 4 & 53 $\pm$ 5 & 57 $\pm$ 3 & 62 $\pm$ 6 & ...\\ 
$\zeta$ Gem   & 60$^{+25}_{-14}$ & 65 $\pm$ 12 & ... & 86 $\pm$ 4 & ... & 68 $\pm$ 3 \\
\enddata

\tablecomments{All measurements are in units of R$_{\odot}$.
There is no previously published estimate for the radius of Polaris.}
\tablenotetext{a}{Surface brightness technique, Moffett and Barnes 1987.}
\tablenotetext{b}{Infrared flux method, Fernley et al. 1989.}
\tablenotetext{c}{CORS method, Ripepi et al. 1997, Caccin et al. 1981.}
\tablenotetext{d}{Infrared spectroscopy and SB technique, Sasselov \& Lester 1990.}
\tablenotetext{e}{Optical Baade-Wesselink method, Scarfe 1976.}
\end{deluxetable}

For Polaris, it has been observed that
the amplitude of the photometric and radial velocity
variations has decreased steadily \citep{fer83,kel84}
and although there is indication that this decrease has
stopped, the amplitudes of these variations
are currently at the level of only 0.032 mag and $\sim$1.7 km s$^{-1}$ \citep{kaf98}.
A surface brightness analysis based on such small
variations is impractical, the radius change would be on the order of 
0.2 R$_{\odot}$, representing an angular diameter change less than 0.5\%.
As a result, there are no published radius estimates
with which we may make a comparison.
For an evaluation of the accuracy of the NPOI linear radius for Polaris
we make a comparison in the following section to
various published period-radius relations derived from both theory and the
application of surface brightness methods to large samples of cepheids.

\section{PERIOD-RADIUS RELATIONS AND POLARIS}

Once a star is identified as a cepheid, the pulsation period is the
one quantity that is always known.
Period-radius relations (hereafter P-R) are therefore powerful tools
for determining the radius of
even the most distant cepheid. Typically, P-R relations are
of the form:

\begin{equation}
\log R = a + b \log P
\end{equation}

\noindent
where $R$ is the radius in units of solar radii, $P$ is the
period in days, and $a$ and $b$ are determined through
observation of cepheids for which the radius can be estimated.
Different methods of determining cepheid radii have in the
past tended to yield somewhat different P-R relations \citep{frn84,
mab87}. We present here a few representative methods from
the literature. Table 3 lists the
derived $a$ and $b$ coefficients for each method.

\begin{deluxetable}{lrrcc}
\footnotesize
\tablecaption{Cepheid Period-Radius Relations\label{tbl-3}}
\tablewidth{0pt}
\tablehead{
\colhead{Method} & \colhead{a}   & \colhead{b}   &
\colhead{R/R$_{\odot}$\tablenotemark{a}} & \colhead{Reference}
}
\startdata
Theory & 1.188 $\pm$ 0.008 & 0.655 $\pm$ 0.006 & 47.6 $\pm$ 1.0 & 1 \\
Opt. SB& 1.146 $\pm$ 0.025 & 0.680 $\pm$ 0.017 & 45 $\pm$ 3\tablenotemark{b} & 2 \\
IR SB  & 1.070 $\pm$ 0.027 & 0.751 $\pm$ 0.026 & 43 $\pm$ 3 & 3 \\
\enddata

\tablenotetext{a}{For a Period = 5.5957 days, logP = 0.7478.}
\tablenotetext{b}{Does not include their intrinsic width of the P-R relation.}
\tablerefs{(1) Bono et al. 1998; (2) Gieren et al. 1999; (3) Laney \& Stobie 1995.}
\end{deluxetable}

\citet{bon98} calculate theoretical P-R relations using
full-amplitude, nonlinear convective models for a variety
of metallicities and stellar masses. The coefficients
for a metallicity
representing Galactic cepheids (Z = 0.02) are given
in Table 3.

\citet{gmb99} use the Baade-Wesselink (or surface brightness)
technique employing
V and V-R photometry, with calibration by \citet{fag97}, to derive
the radii of 116 cepheids in both the Galaxy and Magellanic clouds.
They find no evidence for a difference between Galactic and Magellanic
relations and so calculate a single relation for both.
In addition, \citet{gmb99} find an
intrinsic width to their P-R relation of $\log R \pm 0.03$
which allows for radii consistent with \citet{bon98} over the
range of periods in this paper.

While still using the surface brightness technique for
estimating cepheid diameters, \citet{las95} find K and
J-K (as well as V-K) photometry yield more accurate
results than optical photometry, due to minimal
effects of gravity and microturbulence on
infrared fluxes. For periods less than 11.8 days, \citet{las95}
derive smaller radii than the other two methods.

For periods less than 48 days, which is the range within
which all of the cepheids in this paper are found, the theoretical relation
predicts a larger radius ($\sim$7\%) than that found from surface
brightness relations.
These representative P-R relations are shown in
Figure 2 along with the 
four cepheid radii measured at the NPOI as given in Table 1.
Figure 2 shows that although Polaris has the
highest radius precision (owing to the most precise parallax)
its radius is larger than predicted by any of the published P-R
relations. The difference between the theoretical curve
of \citet{bon98}, the relation which predicts the largest
radius, and the measured radius of Polaris is 2.6$\sigma$ given
the uncertainty of
only 3$R_{\odot}$ in $R_N$.
Even with the intrinsic width of the \citet{gmb99} 
P-R relation, the observed radius for Polaris is too large.
This problem is resolved if Polaris is a first overtone
pulsator rather than a fundamental mode pulsator.
Since the ratio of the first overtone
period to the fundamental period is 0.71, an overtone
cepheid plotted on a P-R diagram using the log of the first overtone
period instead of the fundamental period, will result in a 
radius larger than what the P-R relation would predict
\citep{gbm89}.

\begin{figure}
\figurenum{2}
\plotone{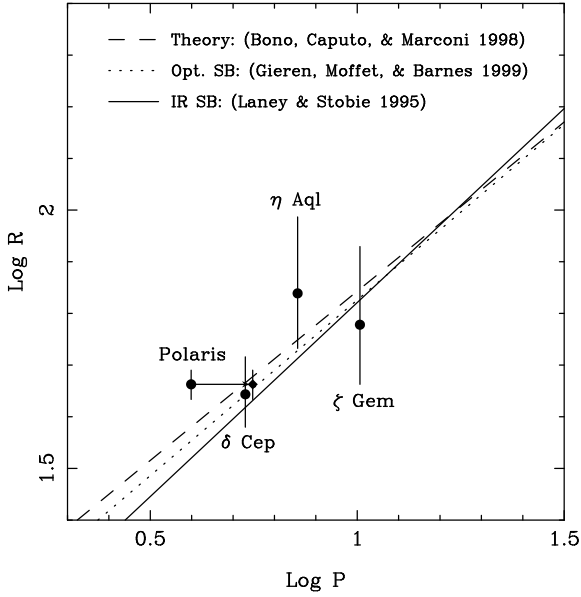}
\figcaption[fig2.ps]{Period-radius diagram for the four observed
cepheids. Shown for comparison is the theoretical relation for
Galactic cepheids \citep{bon98}, the \citet{gmb99} optical
surface brightness relation, and the IR surface brightness relation
derived by \citet{las95}. The circular data point for
Polaris is plotted at the observed period of 3.97 days
while the diamond is the
radius of Polaris plotted at the derived fundamental period
of 5.59 days.}
\end{figure}

The overtone nature of Polaris has been noted recently
in the literature \citep{cox98, fac97}.
\citet{fac97} first used
Hipparcos parallaxes
and visual magnitudes for 220
cepheids to calculate the cepheid period-luminosity zero-point. They
find that the zero-point derived from Polaris alone is brighter
than that produced by the rest of the sample
if Polaris is considered as a fundamental pulsator.
The arrow and diamond in Figure 2 places the measured radius of
Polaris relative to
its fundamental period: 3.9729/0.71 = 5.5957 days. Column 4 of Table 3
lists the radius predicted by each P-R relation for a P = 5.5957 day cepheid.
The percent difference
between the observed radius of Polaris (46 $\pm$ 3 R$_{\odot}$)
and the fundamental period radius predicted by
\citet{bon98} and \citet{gmb99} is 3\% and
2\% respectively.
The excellent agreement of Polaris
with these radii derived from published P-R relations
is evidence of the
overtone nature of Polaris suggested by \citet{fac97}.

\section{CEPHEID MASSES AND $\eta$ AQL}

In the same way that P-R relations yield radii from a known period,
there are period-mass and period-radius-mass relations from
which masses can be derived. The period-radius-mass relation
of \citet{fri72},

\begin{equation}
P = 0.025(M/M_{\odot})^{-0.67}(R/R_{\odot})^{1.70} (days),
\end{equation}

\noindent when solved for mass yields

\begin{equation}
M/M_{\odot} = 4.1 \times 10^{-3}(R_N/R_{\odot})^{2.54}P^{-1.49}.
\end{equation}

\noindent From this equation \citet{gie89}
calculates masses for a sample of 101 cepheids using the
radii of \citet{mab87}.
These masses, which we refer to as $M_{GMB}$ in this paper,
are used by \citet{gie89} to derive the
radius independent Wesselink period-mass (hereafter P-M) relation:

\begin{equation}
M_{WES}/M_{\odot} = 6.30 - 6.06 \log P + 6.28(\log P)^2.
\end{equation}

\noindent Using equation 3 and the radii in Table 1
we calculate masses, $M_N$, for the four cepheids in our sample.
These masses are presented in Table 4 with masses for the four
cepheids from \citet{gie89}.

\begin{deluxetable}{lrcccc}
\footnotesize
\tablecaption{Cepheid Masses Comparison\label{tbl-4}}
\tablewidth{0pt}
\tablehead{
\colhead{Cepheid} & \colhead{$M_{N}$}   & \colhead{$M_{GMB}$\tablenotemark{a}}   &
\colhead{$M_{WES}$} & \colhead{$M_{EV}$} & \colhead{$M_{PUL}$}
}
\startdata
Polaris\tablenotemark{b}  & 5.3$^{+0.9}_{-0.9}$  & ...         & 5.3 $\pm$ 0.9 & 5.8 $\pm$ 0.5 & 4.5 $\pm$ 2.0\\
$\delta$ Cep  & 5.0$^{+2.3}_{-1.7}$  & 3.9 & 5.2 $\pm$ 0.9 & 5.7 $\pm$ 0.5 & 4.5 $\pm$ 2.0\\
$\eta$ Aql    & 10$^{+10}_{-6}$ & 5.1 & 5.7 $\pm$ 0.9 & 6.2 $\pm$ 0.5 & 5.0 $\pm$ 2.0\\ 
$\zeta$ Gem   & 4.2$^{+4.4}_{-2.5}$  & 4.7 & 6.6 $\pm$ 0.9 & 7.0 $\pm$ 0.5 & 6.0 $\pm$ 2.0\\
\enddata

\tablenotetext{a}{Gieren 1989 using surface brightness
radii of Moffett \& Barnes 1987}
\tablenotetext{b}{Calculated at the fundamental mode, P = 5.5957 days.}
\end{deluxetable}

As can be seen from equation 3 and Table 4, large
uncertainties in the linear radius propagate into even larger
uncertainties in the mass, resulting in errors
of almost 50\% for $\delta$ Cep and
100\% towards larger masses for $\eta$ Aql. Compare this to Polaris which has
the highest radius precision and thus
a mass uncertainty of only 17\%. This percent error is
slightly less than the 20\% ``accidental'' error \citet{gie89}
estimates for $M_{GMB}$ based on errors in their radii of
$\sim$7-8\%.

Masses from two theoretical P-M relations are also listed in Table 4: evolution
mass and pulsation mass \citep{gie89}.
The evolution mass, $M_{EV}$, is calculated from
stellar evolution theory:

\begin{equation}
M_{EV}/M_{\odot} = 4.90 - 1.46 \log P + 3.55(\log P)^2
\end{equation}

\noindent whereas the pulsation mass, $M_{PUL}$, is calculated from
period-effective temperature relations:

\begin{equation}
M_{PUL}/M_{\odot} = 5.39 - 6.08 \log P + 6.60(\log P)^2.
\end{equation}

Figure 3 shows the three P-M relations relative to
$M_N$.
As with the estimated error of $M_{GMB}$ the
uncertainty in $M_{EV}$ is estimated to be on the
order of 15-20\%, while uncertainties in the effective
temperature scale for cepheids are capable of bringing
the pulsation mass relation into agreement
with the evolution mass \citep{gie89}. Direct diameter
measurements will be able to address the uncertainties
in this scale once one can
directly measure cepheid diameter
variations and thus calculate effective temperature
as a function of pulsation phase 
to a precision limited only by the photometry. Within the
present uncertainties then, all three P-M relations
are consistent with each other and the NPOI masses.

\begin{figure}
\figurenum{3}
\plotone{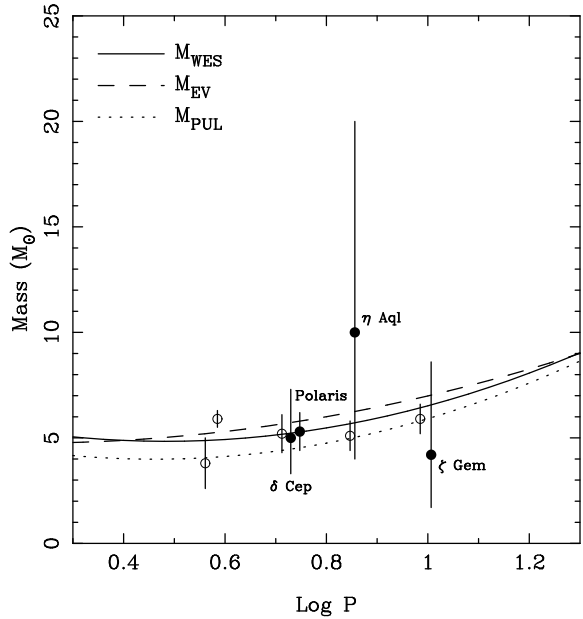}
\figcaption[fig3.ps]{Period-mass diagram for the four cepheids (solid circles).
Shown are three P-M relations from the literature: Wesselink mass,
$M_{WES}$, evolution mass, $M_{EV}$, and pulsation mass, $M_{PUL}$.
Polaris is plotted at the fundamental period. The open circles are the
binary cepheid masses of \citet{eva98}.}
\end{figure}

The cepheid $\eta$ Aql, with a period $\sim$7 days,
possesses a bump in the descending phase of its
radial velocity and light curves. Radii for bump
cepheids (6 days $<$ P $<$ 20 days) can be calculated
using the bump phase and pulsation theory \citep{gbm89,
frn84}. There is a discrepancy, however, between
the radius derived in this manner and the radius
(and hence mass) derived from surface brightness techniques
\citep{gie89}. For a cepheid
such as $\eta$ Aql, $M_{GMB}$ will be larger than the
predicted bump mass. Since the linear radius observed 
by the NPOI for $\eta$ Aql is larger than derived
by \citet{gbm89}, $M_N$ will be larger than $M_{GMB}$ and
therefore in even worse agreement with the mass from the predicted
bump phase method (although the error
in $M_N$ is quite large owing to the large parallax uncertainty).

Recent theoretical models of bump cepheids by \citet{bon00} using full
amplitude, nonlinear, convective models (with no convective
core overshooting) result in a cepheid mass of 6.9 $\pm$ 0.9 M$_{\odot}$
for a period of 11.2 days. This result agrees well
with the Wesselink mass of Equation (4) for a 11.2 day cepheid which
yields M$_{WES}$ = 6.85 M$_{\odot}$.
Although \citet{bon00}
calculate the mass, and thus make their comparison to
\citet{gie89} for a P = 11.2 day cepheid, the implication
is that the disagreement between observation and theory
has been resolved in the matter of bump cepheid masses
with a resolution in favor of the larger Wesselink
mass, and thus closer to $M_N$ in Table 4.

The agreement between Polaris and $\delta$ Cep and
the curve for $M_{WES}$ is less significant than
it would at first seem since the method for
calculating $M_N$ is based upon the same theory as
that used for calculating $M_{WES}$.
In the same way that it is desirable to
compare the model-independent cepheid radii of Table 1 to radii derived
through indirect methods, the same should be done for mass. A number
of cepheids are located in binary systems (including all four cepheids
in this study as noted earlier).
Already \citet{eva98} have used spectroscopy to calculate the
masses of five cepheids in binaries: U Aql, S Mus, V350 Sgr, Y Car, SU Cyg.
Using ground based optical and satellite ultraviolet spectra
the mass ratio of the two members of the binary were found.
Inferring the mass of the companion based on the spectral type
yielded the mass of the cepheid. These five
cepheids are plotted in Figure 3 as open circles and show good agreement
with the P-M relations and the NPOI observations.
With long enough
baselines, optical interferometry will be able to
image the orbits of cepheids as has already been
done for binaries of non-cepheids
\citep{ben97,hum98}. In conjunction
with radial velocities from
spectroscopy, all the orbital elements of the system,
including the mass and distance, will be directly determined and
independent of all models.

\section{FUTURE OBSERVATIONS}

At the present time only four cepheids have had their
diameters measured with the NPOI. Over the next two
years as the longest baseline available increases
from 38 meters to 440 meters, the number of cepheids
resolvable and the precision of their measurements
will increase by at least a factor of five. Figure 2
shows that at the present the measured
linear radii are consistent with each of the published
P-R relations. With the increased
precision of the angular diameter measurements, and
the increased precision of parallax observations (also
undertaken at the USNO) it
will be possible to differentiate between the various
P-R relations which are seen to diverge in Figure 2
for periods shorter than 30 days. This is precisely the
range of periods for which cepheids observable by the
NPOI are located.

\section{SUMMARY}

Optical long baseline interferometry has successfully measured
the mean angular diameters of the four brightest cepheid
variables in the northern sky. These angular diameters coupled
with trigonometric parallaxes have produced virtually model-independent
linear radii. These radii are compared to radii
in the literature which have been derived from a variety
of Baade-Wesselink, or surface brightness, methods.
The agreement between the direct radius determinations presented here
and published indirect radius estimates
is quite good. The differences are
$\sim$10\%, better than the error in $R_N$ which is on the order of 20-40\%.
For $\eta$ Aql, the derived linear radius is in marginal agreement
with the optical surface brightness results but it is in
very good agreement with the radius estimated from
infrared spectroscopy by \citet{sas90}.
For Polaris the radius precision (6\%) is high enough that we
are able to confirm its overtone nature. At a period of 3.97 days
a radius of 46 $\pm$ 3 R$_{\odot}$ is inconsistent with
the published P-R relations of \citet{bon98} and \citet{gmb99}.
Only as an overtone pulsator with a fundamental period of 5.59 days
is Polaris in agreement with these P-R relations, confirming the findings
of \citet{fac97} using completely independent
means.
At a period of 5.59 days,
the Wesselink mass of Polaris is found to be in excellent agreement
with the period-mass relation of \citet{gie89}.

\begin{acknowledgments}
The authors would like to thank Hugh Harris for providing
results from the USNO variable star parallax study.
Thanks are also due to Siobahn Morgan for bringing the Polaris overtone
pulsator discussion to our attention. Don Fernie was also enormously
helpful in providing the period and rate of change for Polaris.
\end{acknowledgments}


\begin{thebibliography}{}


\bibitem[Armstrong et al.(2000)]{arm00} Armstrong, J. T., Nordgren, T. E., Mozurkewich,
Germain, M. E., Hajian, A. R., Hindsley, R. B., Hummel, C. A., \& Thessin, R. N.
2000, \aj, in press

\bibitem[Barnes \& Evans(1976)]{bae76}Barnes, T. G. \& Evans, D. S. 1976,
\mnras, 174, 489


\bibitem[Benson et al.(1997)]{ben97}Benson, J. A., Hutter, D. J., Elias~II,
N. M., Bowers, P. F., Johnston, K. J., Hajian, A. R., Armstrong, J. T.,
Mozurkewich, D., Pauls, T. A., Rickard, L. J, Hummel, C. A., White, N. M.,
Black, D., \& Denison, C. S. 1997, \aj, 114, 1221


\bibitem[B\"{o}hm-Vitense \& Proffitt(1985)]{vap85}B\"{o}hm-Vitense, E.
 \& Proffitt, C. 1985, \apj, 296, 175

\bibitem[Bono, Caputo, \& Marconi(1998)]{bon98}Bono, G., Caputo, F., \& Marconi, M. 1998,
\apj, 497, L43

\bibitem[Bono, Marconi, \& Stellingwerf(2000)]{bon00}Bono, G., Marconi, M., \& Stellingwerf, R. F. 2000,
in press

\bibitem[Caccin et al.(1981)]{cor81}Caccin, B., Onnembo, A., Russo, G., \& Sollazzo, C. 1981, \aap, 97, 104

\bibitem[Claret, Dias-Cordova, \& Gimenez(1995)]{cla95}Claret, A., Dias-Cordova, J.,
\& Gimenez, A. 1995, Astronomy and Astrophysics Supplement, 114, 247

\bibitem[Cox(1998)]{cox98}Cox, A. N. 1998, \apj, 496, 246

\bibitem[Evans et al.(1998)]{eva98}Evans, N. R., B\"{o}hm-Vitense, E., Carpenter, K., Beck-Winchatz, B.,
\& Robinson, R. 1998, \apj, 494, 768

\bibitem[Feast \& Catchpole(1997)]{fac97}Feast, M. W., \& Catchpole, R. M.
1997, \mnras, 286, L1

\bibitem[Fernie(1984)]{frn84}Fernie, J. D. 1984, \apj, 282, 641

\bibitem[Ferro(1983)]{fer83}Ferro, A. A. 1983, \apj, 274, 755

\bibitem[Fernley, Skillen, \& Jameson(1989)]{fly89}Fernley, J. A., Skillen, I., \& Jameson,
R. F. 1989, \mnras, 237, 947

\bibitem[Fouqu\'{e} \& Gieren(1997)]{fag97}Fouqu\'{e}, P. \& Gieren, W. P.
1997, \aap, 320, 799

\bibitem[Fricke, Stobie, \& Strittmatter(1972)]{fri72}Fricke, K., Stobie, R. S., \& Strittmatter,
P. A. 1972, \apj, 171, 593

\bibitem[Gieren(1989)]{gie89}Gieren, W. P. 1989, \aap, 225, 381

\bibitem[Gieren, Barnes, \& Moffett(1989)]{gbm89}Gieren, W. P., Barnes~III, T. G., \&
Moffett, T. J. 1989, \apj, 342, 467

\bibitem[Gieren, Moffett, \& Barnes(1999)]{gmb99}Gieren, W. P., Moffett, T. J., \&
Barnes~III, T. G. 1999, \apj, 512, 553

\bibitem[Gieren et al.(2000)]{gie00}Gieren, W. P., Storm, J., Fouqu\'{e}, P.,
Mennickent, R. E., \& Gomez, M. 2000, \apjl, 533, 107

\bibitem[Hajian et al.(1998)]{haj98}Hajian, A. R., Armstrong, J. T.,
Hummel, C. A., Benson, J. A., Mozurkewich, D.,
Pauls, T. A., Hutter, D. J., Elias II, N. M., Johnston, K. J.,
Rickard, L. J, \& White, N. M. 1998, \apj, 496, 484


\bibitem[Hoffleit \& Jaschek(1982)]{hof82}Hoffleit, D., \& Jaschek, C. 1982,
The Bright Star Catalogue (4th rev. ed.; New Haven: Yale Univ. Obs.)

\bibitem[Hummel et al.(1998)]{hum98}Hummel, C. A., Mozurkewich, D.,
Armstrong, J. T., Hajian, A. R., Elias II, N. M., \& Hutter, D. J. 1998, AJ,
116, 2536

\bibitem[Kamper, Evans, \& Lyons(1984)]{kel84}Kamper, K. W., Evans, N. R., \& Lyons, R. W.
1984, \jrasc, 78, 173

\bibitem[Kamper \& Fernie(1998)]{kaf98}Kamper, K. W. \& Fernie, J. D. 1998, \aj, 116, 936

\bibitem[Laney \& Stobie(1995)]{las95}Laney, C. D. \& Stobie, R. S.
1995, \mnras, 274, 337

\bibitem[Moffett \& Barnes(1987)]{mab87}Moffett, T. J., \&
Barnes~III, T. G. 1987, \apj, 323, 280



\bibitem[Nordgren et al.(1999)]{ten99}Nordgren, T. E., Germain, M. E., 
Benson, J. A., Mozurkewich, D., Sudol, J. J., Elias II, N. M., 
Hajian, A. R., White, N. M., Hutter, D. J., Johnston, K. J.,
Gauss, F. S., Armstrong, J. T., Pauls, T. A., \& Rickard, L. J 1999, \aj,
118, 3032

\bibitem[Perryman et al.(1997)]{per97}Perryman, M. A. C., Lindegren,
L., Kovalevsky, J., H\"{o}g, E., Bastian, U.,
Bernacca, P. L., Creze, M., Donati, F., Grenon, M., Grewing, M.,
Van Leeuwen, F., Van der Marelh, H., Mignard, F., Murray, C. A.,
Le Poole, R. S., Schrijver, H., Turon, C., Arenou, F., Froeschle,
M., \& Petersen, C. S. 1997, \aap, 323, L49


\bibitem[Ripepi et al.(1997)]{rip97}Ripepi, V., Barone, F., Milano, L., \& Russo, G.
1997, \aap, 318, 797


\bibitem[Sasselov \& Lester(1990)]{sas90}Sasselov, D. R., \& Lester, J. B.
1990, \apj, 362, 333

\bibitem[Scarfe(1976)]{sca76}Scarfe, C. 1976, \apj, 209, 141

\bibitem[Szabados(1980)]{sza80}Szabados, , L. 1980, Mitteilungen der Sternwarte
der ungarischen Akademie der Wissenschaften, Budapest, 76, 32

     
\bibitem[Straizys \& Kuriliene(1981)]{str81}Straizys, V. \& Kuriliene, G.
1981, \apss, 80, 353

\bibitem[Wesselink(1969)]{wes69}Wesselink, A. J. 1969, \mnras, 144, 297

\end{thebibliography}
\end{document}